# Monte Carlo Simulation for Particle Detectors

Maria Grazia Pia[1] and Georg Weidenspointner[2,3] *on behalf of the NANO5 Team*

[1]INFN Sezione di Genova, Italy
[2]Max-Planck-Institut Halbleiterlabor, München, Germany
[3]Max-Planck-Institut für extraterrestrische Physik, Garching, Germany

## 1. Overview of current status and future perspectives

Monte Carlo simulation is an essential component of experimental particle physics in all the phases of its life-cycle: the investigation of the physics reach of detector concepts, the design of facilities and detectors, the development and optimization of data reconstruction software, the data analysis for the production of physics results.

This major role is objectively documented by scientometric data: the main reference of the Geant4 simulation toolkit [1] has achieved more than 3000 citations (3122 at the time of writing this note) [2], and is the most cited article authored by CERN over the time frame covered by Thomson-Reuters' database [2] (since 1970). It is worthwhile to note that the three top cited papers [1][3][4] in Thomson-Reuters' Instruments & Instrumentation category concern particle transport simulation. Although Geant4 was originally motivated by the simulation requirements of LHC experiments, the analysis of its citations [5] shows that it is widely used in various other physics research environments – astrophysics, nuclear physics and engineering, medical physics etc., and in industry.

This note briefly outlines some research topics related to Monte Carlo simulation, that are relevant to future experimental perspectives in particle physics. The focus is on physics aspects: the experience of the past decades demonstrates that this simulation domain requires wide discussion and cooperation among various components of the research community.

The authors of this note are actively engaged in R&D for future simulation methods; a bibliography of related developments and results is available in [6].

## 2. Physics Capabilities

Extended physics modeling capabilities are needed to effectively support the R&D for future detectors and facilities. Some areas requiring investment are briefly outlined below; further discussions in the experimental community are encouraged to identify current and future physics modeling requirements. A dedicated workshop, or a series of dedicated regular meetings for this purpose would be beneficial.

The development and validation of simulation capabilities for future detector R&D requires the investment of dedicated resources. It is worthwhile to note that this is a complex research domain, which requires a variety of competence – in physics, mathematical methods, software technology and epistemology.

## 1. State of the art of physics modeling

The assessment of the state-of-the-art in physics models and parameters relevant to particle transport, and their implementation in the software, contributes to the reliability of the simulation outcome. Recent studies have highlighted that physics models and parameters implemented in major Monte

Carlo systems do not always reflect the body of knowledge established in recent years; efforts invested in this area result in significantly improved simulation accuracy and quantified reliability [7]-[9].

The investment to embody the state of the art in the physics scope of Monte Carlo codes is propaedeutic to further extensions; it should be pursued with adequate resources.

## 2. Physics models for future generation's detectors

Current Monte Carlo systems are based on the so-called "independent particle approximation" (IPA): this scheme describes adequately only a limited set of physics processes, which substantially correspond to the experimental needs of detector modeling in the 70's or 80's of the past century. The adoption of the IPA scheme implies that materials are approximated in the simulation as mixtures of single atom gases. This very basic assumption breaks down when one intends to simulate detector features such as particle transport in solids, scintillator non-proportionality, the exploitation of nanotechnologies, and more in general detailed effects related to material properties, where dealing with molecules or solids, rather than with gases, makes a difference. The limitations of IPA are already inadequate to the simulation of current experimental technologies, and are expected to become even more relevant in the future.

An investment to incorporate the body of knowledge of materials science relevant to novel detection technologies is needed to effectively support next generation's detectors and experiments.

It is worthwhile to note that moving away from the IPA scheme would require not only appropriate models of particle interactions with materials, but also significant rethinking of the particle transport schemes that have been the basis for detector simulation for the past fifty years. The necessary conceptual and technical investment for such a major evolution in particle transport should not be underestimated.

## 3. Radiation damage

Current Monte Carlo systems have limited capabilities to predict the damage to detectors and electronic components exposed to radiation. The availability of a simulation system with reliable predictive capabilities in this domain would significantly support the design of future detectors.

To a large extent, the response to this issue falls into the topic discussed in the previous section.

## 4. Multi-scale simulation

Current Monte Carlo systems adopt mixed condensed-random-walk and discrete transport schemes to deal with infrared divergence affecting some particle interaction processes (e.g. ionization), since detailed simulation of soft secondaries would require prohibitively extensive computational resources. Although conceptually appealing and appropriate to many simulation applications, this scheme suffers from drawbacks [10]; it fails when multi-scale simulation is needed to model detector behaviour, such as when the transition from a nano-scale structure (e.g. quantum dots, nanowires etc.) to a macroscopic system (e.g. a scintillator) is involved.

Exploratory investigations of techniques supporting multi-scale simulation [11] are in progress; further investment is needed to establish this functionality in general purpose Monte Carlo systems.

# 3. Validation and Uncertainty Quantification

The experimental validation of the physics models implemented in the simulation, and the quantitative estimate of related uncertainties, is essential for the reliability of Monte Carlo simulation as a predictive experimental instrument.

Only a relatively small fraction of the basic physics constituents of current Monte Carlo codes – cross sections, angular distributions, secondary spectra, atomic parameters etc. – is documented in the literature as having been quantitatively validated.

Often the validation of a simulation system is limited to qualitatively assessing its capability to reproduce some macroscopic features of an experimental scenario: this gross validation is inadequate to determine the predictive power of the simulation. The lack of in-depth validation of all the simulation components often results in inability of the simulation system to describe detailed features, such as the tails in distributions of detector observables. Such a level of detail may be important in detector design and physics analysis. It is not uncommon that the lack of thorough and rigorous validation of all simulation components may affect not only detailed features, but also major observable patterns.

The validation of physics models implemented in Monte Carlo codes is often hindered by the lack of pertinent experimental data, or their poor quality (e.g. conflicting measurements, evidence of systematic effects, undocumented or poorly estimated experimental uncertainties etc.).

An effort should be invested in the quantitative validation of physics models implemented in Monte Carlo codes. If necessary, one should envisage dedicated experiments for the validation of simulation models [12].

The quantification of simulation uncertainties, and of how uncertainties in physics quantities embedded in simulation models affect the outcome of the simulation, has been object of study in the context of deterministic simulation, while it is at the very beginning in the domain of Monte Carlo simulation. An investment in this domain is propaedeutic to determine objectively the predictive power of simulation systems.

Special attention should be devoted to epistemic uncertainties, i.e. uncertainties due to intrinsic lack of knowledge, affecting physics models implemented in simulation systems. Epistemic uncertainties are frequently embedded in hadronic models, and are often not documented, i.e. simulation users may not be aware of their presence. The identification and quantification of epistemic uncertainties, and of the systematic effects they may induce in experimental observables, is critical to determine the reliability of the simulation [13]. Adequate methods for their treatment in Monte Carlo simulation should be developed.

An investment is needed in parallel to identify statistical analysis methods appropriate to the validation of simulation systems, and to develop related software tools. Preliminary empirical observations have highlighted the inadequacy of conventional goodness-of-fit tests in some experimental scenarios [7] [9]; limited documentation of their power is available in the literature.

## 4. Data Libraries

Data libraries, i.e. tabulations of the results of complex theoretical calculations or of evaluated experimental data, are an important aid for physics modeling in Monte Carlo simulation systems. Their development and open distribution should be promoted among theoretical and experimental groups. Data distribution centers, such as NEA and RSICC, already exist for the distribution of data libraries; cooperation with these centers would be desirable.

The availability of open documentation and data for experimental benchmarks would facilitate the validation of simulation models. An appropriate infrastructure to manage open experimental benchmark material (and results) should be promoted and supported.

## 5. Cooperation and Synergy

By its own nature, Monte Carlo simulation for particle transport encompasses in-depth knowledge of various scientific domains. Cooperation of detector experts, physics simulation experts, theoreticians, atomic and nuclear physicists, mathematicians and statisticians, and software engineers is needed to respond effectively to the needs of future experimental R&D.

Many of the problems faced in R&D for particle physics experiments are common to other experimental domains, such as astrophysics, photon science, nuclear physics and medical physics.

Synergy across different experimental communities should be promoted to profit from more extensive expertise and to share the investment of resources for common simulation tools.

Dedicated, low cost workshops should be promoted to facilitate communication and cooperation for the development and validation of detector simulation.

Special attention should be devoted to the education of new generations of simulation experts in the particle physics community and the support of their academic careers: this aspect is often neglected, with the result that expertise in this critical experimental domain is scarce, and often cannot be transferred to younger researchers.

## 6. Additional Benefits

The widespread use of Geant4 demonstrates the impact of detector simulation on multidisciplinary experimental domains. Further investments in simulation R&D motivated by particle physics requirements are expected to have significant impact; they would offer opportunities for technology transfer to industry, and would benefit the civil society. If properly communicated to the media, the social impact of simulation tools would contribute to the image of particle physics research.

## 7. Conclusions

R&D is needed to produce and validate physics simulation tools capable of modeling future detectors and experimental facilities. Pioneering projects in various areas – physics models, techniques for multi-scale simulation, quantitative validation and uncertainty quantification – have demonstrated preliminary concrete achievements.

Wide cooperation within the particle physics community, and with other scientific domains sharing similar requirements, should be pursued, and adequate resources should be invested, to develop quantitatively validated simulation tools for future experiments.

## References

[1]   S. Agostinelli et al., "Geant4 - a Simulation Toolkit", *Nucl. Instrum. Meth. A*, vol. 506, pp. 250-303, 2003.
[2]   Thomson-Reuters, ISI Web of Knowledge℠, http://apps.webofknowledge.com.
[3]   J. P. Biersack and L. G. Haggmark, "A Monte-Carlo computer-program for the transport of energetic ions in amorphous targets", *Nucl. Instrum. Meth.*, vol.174, no. 1-2, pp. 257-269, 1980.
[4]   L. R. Doolittle, "Algorithms for the rapid simulation of Rutherford backscattering spectra", *Nucl. Instrum. Meth. B*, vol. 9, no. 3, pp. 344-351, 1985.
[5]   M. G. Pia, T. Basaglia, Z. W. Bell, P. V. Dressendorfer, "Geant4 in scientific literature," *Proc. IEEE Nucl. Sci. Symp.*, pp. 189-194, 2009.
[6]   Bibliography: http://www.ge.infn.it/geant4/papers/.
[7]   M. Batic et al., "Photon elastic scattering simulation: validation and improvements to Geant4", *IEEE Trans. Nucl. Sci.*, vol. 59, no. 64, Aug. 2012, doi: 10.1109/TNS.2012.2203609 , arXiv:1206.0498 preprint.
[8]   H. Seo et al., "Ionization cross sections for low energy electron transport", *IEEE Trans. Nucl. Sci.*, vol. 58, no. 6, 2011.
[9]   M. G. Pia et al., "Evaluation of atomic electron binding energies for Monte Carlo particle transport", *IEEE Trans. Nucl. Sci.*, vol. 58, no. 6, Dec. 2011
[10]  M. G. Pia, G. Weidenspointner, M. Augelli, L. Quintieri, P. Saracco, M. Sudhakar, A. Zoglauer, "PIXE simulation with Geant4", *IEEE Trans. Nucl. Sci.*, vol. 56, no. 6, pp. 3614-3649, 2009.
[11]  M. Augelli et al., "Environmental Adaptability and Mutants: Exploring New Concepts in Particle Transport for Multi-Scale Simulation", *Proc. IEEE Nucl. Sci. Symp.*, pp. 1153-1154, 2010.
[12]  D. E. Post and L. G. Votta, "Computational Science Demands a New Paradigm", Phys. Today, pp. 35-41, 2005.
[13]  M. G. Pia, M. Begalli, A. Lechner, L. Quintieri, P. Saracco, "Physics-related epistemic uncertainties of proton depth dose simulation", *IEEE Trans. Nucl. Sci.*, vol. 57, no. 5, pp. 2805-2830, 2010.